\documentclass[a4paper,dvips,12pt]{article}
\usepackage{epsfig,graphics,graphicx}
\begin{document}
\pagestyle{myheadings} \markright{\it   CBPF-NF-030/02} \vskip.5in
\begin{center}
%
%
\vskip.4in {\Large\bf DIRAC OSCILLATOR VIA R-DEFORMED HEISENBERG ALGEBRA }
\vskip.3in
%
%
%
R. de Lima Rodrigues\footnote{Permanent address:
Departamento de Ci\^encias
Exatas e da Natureza, Universidade Federal de Campina Grande,
Cajazeiras - PB, 58.900-000-Brazil. \tt E-mail to RLR is
rafaelr@cbpf.br or rafael@fisica.ufpb.br,
the \tt e-mail to ANV is vaidya@if.ufrj.br.}\\
Centro Brasileiro de Pesquisas F\'\i sicas (CBPF)\\
Rua Dr. Xavier Sigaud, 150, CEP 22290-180, Rio de Janeiro-RJ, Brazil\\
A. N. Vaidya\\
Instituto de F\'\i sica - Universidade Federal do Rio de Janeiro\\
Caixa Postal 68528 - CEP 21945-970, Rio de Janeiro, Brazil
%
%
%
\end{center}

\vskip.2in
\begin{abstract}
The complete energy spectrum for the Dirac
oscillator via R-deformed Heisenberg  algebra
is investigated.
\end{abstract}

\vspace{1cm}

\vspace{1cm}
To appear in the proceedings at the XXIII Brazilian National
Meeting on Particles and Fields, October 15 to 19, 2002, \'Aguas de
Lind\'oia-SP, Brazil, site www.sbf1.if.usp.br/eventos/enfpc/xxiii.

\section{Introduction}

The relativistic Dirac oscillator  proposed by Moshinsky-Szczepaniac
\cite{Mosh89} is a
spin ${1\over 2}$ object with the Hamiltonian which in the
non-relativistic limit  leads to that of a 3-dimensional isotropic oscillator
shifted by a constant term plus a ${\vec L}\cdot {\vec S}$ coupling
term for both signs of energy.
There they construct
a Dirac Hamiltonian, linear in the momentum $\vec p$ and position $\vec r,$
whose square leads to the ordinary harmonic oscillator in the
non-relativistic limit.
The Dirac oscillator have been investigated in several context
\cite{MZ}.

The R-deformed
Heisenberg algebra or Wigner-Heisenberg algebraic
technique \cite{WH} was recently
super-realized for the SUSY isotonic oscillator
\cite{JR1,Mik00}. The R-Heisenberg algebra has also been
investigated for the
three-dimensional non-canonical oscillator to generate a representation
of the  orthosympletic Lie superalgebra $osp(3/2)$ \cite{PS}.

The R-Heisenberg algebra has been
found relevant  in the context of integrable models \cite{Vasiliev91},
and the Calogero interaction  \cite{Poly92,macfa93}.
Recently it has been employed for bosonization of supersymmetry
in quantum mechanics \cite{Mik97},
and the discrete space structure for the 3D Wigner quantum
oscillator has been investigated \cite{palev02}.
In this work, we obtain the complete energy spectrum for the Dirac
oscillator via R-deformed Heisenberg (RDH) algebra.

\section{3D Wigner Oscillator}

In this Section, we provide a three dimension presentation of the
Wigner system with its bosonic sector to be the 3D isotropic oscillator
(assumed to be of spin-$\frac{1}{2}$, to aid factorization).

The R-deformed Heisenberg (or Wigner-Heisenberg)
algebra is given by following (anti-)commutation relations ($[A,B]_+\equiv
AB+BA$ and $[A,B]_-\equiv AB-BA):$

\begin{equation}
\label{RH}
H=\frac 12 [a^-, a^+]_+, \quad
[H, a^{\pm}]_-=\pm a^{\pm}, \quad
 [a^-, a^+]_-=1+c R, \quad
[R, a^{\pm}]_+=0,\quad R^2=1,
\end{equation}
where $c$ is a real constant associated to the Wigner parameter
\cite{JR1}. Note that when $c=0$ we have the
standard Heisenberg algebra.

It is straightforward, following the analogy with the Ref. \cite{JR1},
to define
the super-realizations for the ladder operators $a^{\mp}(\vec \sigma \cdot
\vec L + {\bf 1})$ for $H_W \equiv H(\vec \sigma \cdot \vec L + {\bf 1})$
taking the explicitly forms

\begin{equation}
\label{2-8}
a^{\mp} = a^{\mp}(\vec \sigma \cdot \vec L + {\bf 1}) =
\frac{1}{\sqrt {2}} \left\{\mp \Sigma_1 \left(\frac{\partial}{\partial r} +
\frac{1}{r}\right) \pm  \frac{1}{r}(\vec \sigma \cdot \vec L + {\bf 1})
\Sigma_1 \Sigma_3 - \Sigma_1 r \right\}
\end{equation}
which satisfy together with $H_W \equiv
H(\vec \sigma \cdot \vec L + {\bf 1})$  all
the algebraic relations  of the RDH algebra with the constant $\frac{c}{2}$
replaced by $(\vec \sigma \cdot \vec L + {\bf 1})$ and $R= \Sigma_3.$
Note that $(\vec\sigma\cdot\vec L + {\bf 1})$ commutes with all the
basic elements $(a^{\mp}$ and $H_{W}$) of the RDH algebra.

It may be observed that the RDH algebra that gets defined here
is in fact three
dimensional (one dimension for $r$ and two for $(\vec \sigma \cdot
\vec L + {\bf 1})$) and is identically satisfied on any arbitrary three
dimensional wave function.

On the eigenspaces of the operator $(\vec \sigma \cdot \vec L + {\bf 1})$,
 the 3D Wigner algebra gets reduced to a 1D from with   $(\vec \sigma \cdot
\vec L + {\bf 1})$ replaced by its eigenvalue $\mp(\ell + 1)$,
$\ell=0,1,2,\cdots,$ where $\ell$ is the orbital angular momentum quantum
number. The eigenfuncitons of $(\vec \sigma \cdot \vec L + {\bf 1})$ for
the eigenvalues $(\ell + 1)$ and $-(\ell + 1)$ are respectivaly given
by the well known spin-spherical harmonic $y_{\mp}.$

Now, considering simultaneous eigenfuncitons of the mutually commuting
$H_W$ and $(\vec\sigma\cdot\vec L + {\bf 1})$ by

\begin{equation}
\label{pW+} \psi_{W,+} ={\tilde R_{1,+}(r)\choose\tilde R_{2,+}(r)} y_+,
\quad (\vec \sigma \cdot \vec L + {\bf 1})\psi_{W,+} = (\ell +
1)\psi_{W,+},
\end{equation}

\begin{equation}
\label{2-17}
\psi_{W,-} = {\tilde R_{1,-}(r)\choose\tilde R_{2,-}(r)} y_-, \quad
(\vec \sigma \cdot \vec L + {\bf 1})\psi_{W,-} = -(\ell+1)\psi_{W,-},
\end{equation}
(where the use of the subscript $+(-)$ indicates association with
$[y_+(y_-)$], we observe that the positive semi-definite form of $H_W$ the
ladder relations and the form of $H_W$ dictat that
the ground state energy $E^{(0)}_{w}(\vec \sigma \cdot \vec L + {\bf 1})
\geq 0$, where $E_W(\vec \sigma \cdot \vec L + {\bf 1})$ indicates a
function of $\vec\sigma\cdot\vec L +{\bf 1}$, is determined by the
annihilation condition which reads as
two cases.

\section{The Dirac Oscillator Model}

Adding an "anomalous momentum" in the form of a (nonlocal)
linear and hermitian  interaction,
$
\vec{\alpha }.\vec{\pi } \equiv  -iM\omega \beta \vec{\alpha
}.\vec{r} = (\vec{\alpha }.\vec{\pi })^{\dagger },
$
in the (noncovariant) Dirac free particle equation with
mass $M$ and spin-$\frac 12,$ in the natural sistem of units,

\begin{equation}
\label{ED}
i{\partial \psi \over \partial t} = (\vec{\alpha }.\vec{p} +
M\beta )\psi,
\end{equation}
one obtains the equation for the Dirac oscillator  \cite{Mosh89}:

\begin{equation}
\label{OD}
 i{\partial \psi \over \partial t} = \{\vec{\alpha }.(\vec{p} +
\vec{\pi }) + M\beta \}\psi ,
\end{equation}
where $M$ and $\omega$  are, respectively, the  mass of the  particle and the
frequency of the oscillator, and the  matrices $(\vec{\alpha }, \beta )$
satisfy the following properties:

\begin{equation}
\label{AM}
[\alpha_{i}, \beta ]_{+}= 0, \quad [\alpha _{i}, \alpha _{j}]_{+}=
2\delta_{ij} {\bf 1}, \quad \beta ^{2}=
{\bf 1} = \alpha^{2}_{i}, \quad (i,j=1,2,3).
\end{equation}
Writing the Dirac spinor in terms of the upper and lower components,
respectively, $\psi_{1}$ and $\psi_{2}$,
$
\Psi (\vec{r},t) = \exp (-\hbox{iEt})
\left[\matrix{\psi_{1}(\vec{r}) \cr \psi_{2}(\vec{r})}\right]
$
 the standard representation of the matrices
$\vec{\alpha }$ and $\beta.$

\section{The Dirac oscillator via RDH algebra}

In this section, we implement a new realization of the Dirac oscillator
in terms  of elements of the R-deformed Heisenberg
algebra. To solve the equation Dirac,  following the usual
procedure, we consider the second order differential equation,

\begin{equation}
\label{ENHD}
\tilde{H}_{D}\psi (\vec{r}) = E\psi (\vec{r}),
\end{equation}
where $\tilde{H}_{D}$ is a  second order Hamiltonian,
$
\tilde{H}_{D}= H^{2}_{D} + M^{2}{\bf 1}, \quad\tilde{E} =
{E^{2}- M^{2}\over 2M}.
$
In the spherical polar coordinate system,
we obtain the non-relativistic form of the Hamiltonian Ui \cite{Ui84}, for an
isotropic 3D SUSY harmonic oscillator with spin-$\frac 12.$

We consider a unitary operator
in terms of the radial projection of the spin,

\begin{equation}
\label{MU}
U= \left[\matrix{1&0 \cr 0&\sigma_r}\right]= U^{-1}= U^{\dagger},
\end{equation}
to obtain the following relation between the transformed Dirac Hamiltonian,
$\tilde{H}_{D},$  the 3D Wigner Hamiltonian,
$H_W,$ and the SUSY  Hamiltonian,
$H_{\hbox{SUSI}}$ \cite{Ui84}:

\begin{equation}
\label{HS}
H_{\hbox{SUSI}}=U\tilde{H}_{D}U^{\dagger}=
H_W - {1\over
2}\{1+2(\vec{\sigma\cdot\vec{L}+{\bf 1}})\Sigma_{3}\}\omega \Sigma_{3}.
\end{equation}

\subsection{The energy spectrum of the Dirac oscillator}

The energy spectra of the operators $\tilde{H}_{D}$ and $H_{\hbox{SUSY}}$
are identical, since these operators are related by a unitary
 transformation. However, the relation between the principal
quantum number $N$ and the angular momentum $(\ell)$ is different, in
each case.  Obviously, the energy spectrum
associated with the two types of eigenspaces belonging to the eigenvalues
$\pm (\ell +1)$:

Case(i) $\rightarrow  \vec\sigma\cdot\vec{L}+{\bf 1} \rightarrow  \ell+1 =
j+\frac 12, \quad j = \ell +\frac 12$

\begin{equation}
\label{E+i}
\tilde{E}_{N\ell } = {E^{2}- M^{2}\over 2M} =
\left\{\begin{array}{c}2m\omega  = \tilde{E}^{+}_{N(\ell +1)},
\\
2(m+1)\omega  = \tilde{E}^{-}_{N\ell },
\end{array}\right.
\end{equation}
where $m=0,1,2,\ldots_.$

Case(ii) $\rightarrow  \vec\sigma\cdot\vec{L} +{\bf 1} \rightarrow -(\ell +
1) = -(j + \frac 12), \quad j = (\ell  + 1) -\frac 12:\\
\tilde{E}_{N\ell} = {E^{2}-M^{2}\over 2M}=
\left\{\begin{array}{c}
(N+j+3/2)\omega= \tilde{E}^{+}_{N\ell}, \quad N=j-\frac 12,
j+3/2, j+7/2,\ldots, \\
(N+j+5/2)\omega = \tilde{E}^{-}_{N(\ell +1)},
\quad N=j+\frac 12, j+5/2,\ldots_.
\end{array}\right.$
\section{Conclusion}

In this work we investigate the
Dirac oscillator with the help of techniques
of super-realization of the R-deformed Heisenberg algebra.

The Dirac oscillator with different interactions has been treated by
Casta\~nos {\it et al.}  and by Dixit {\it et al.}
\cite{MZ}. These works
motivate the construction of a new linear Hamiltonian in terms of the
momentum, position and mass coordinates, through a set of seven  mutually
anticommuting 8x8-matrices yielding a representation of the Clifford
algebra $C\ell_7$. The seven elements of the Clifford algebra $C\ell_7$
generate the three linear momentum components, the three position
coordinates components and the mass, and their squares are the 8x8-identity
matrix {\bf I}$_{8\hbox{x}8}.$
Results of
our analysis on Dirac oscillator via  the Clifford algebra $C\ell_7$
are in preparation.

In a forthcoming paper we show that the Dirac oscillator equation can be
resolved algebrically without having to transform it into a second order
diferential equation. Therefore, the important connection for the Dirac
3D-isotropic oscillator with the linear ladder operators of the R-deformed
Heisenberg algebra, satisfying the concomitant general oscillator quantum
rule of Wigner, have explicited in this work.

\vspace{1cm}
\centerline{\bf ACKNOWLEDGMENTS}

The authors are grateful to J. Jayaraman, whose advises and encouragement
were precious. RLR was supported in part by CNPq (Brazilian Research
Agency). He wishes to thank J. A. Helayel Neto for the kind of hospitality
at CBPF-MCT. The authors wish also thank the staff of the CBPF and
DCEN-CFP-UFCG. The authors are also grateful to the organizing commitee of
the of the XXIII Brazilian National Meeting on Particles and Fields
(ENFPC), October 15 to 19, 2002, \'Aguas de Lind\'oia-SP.

\end{document}